\documentclass[twocolumn,preprintnumbers,amsmath,amssymb]{revtex4}

\usepackage{graphicx}
\usepackage{dcolumn}
\usepackage{bm}
\usepackage{multirow}

\def\bra#1{\mathinner{\langle{#1}|}}
\def\ket#1{\mathinner{|{#1}\rangle}}

\def\CZ{\overline{CZ}}
\def\H{\overline{H}}

\begin{document}
\title{Globally controlled quantum wires for perfect qubit transport, mirroring and  quantum computing}

\author{Joseph Fitzsimons }
\email{joe.fitzsimons@materials.ox.ac.uk}
\affiliation{Department of Materials, Oxford University, Oxford, UK}
\author{Jason Twamley}\email{jtwamley@ics.mq.edu.au}
\affiliation{Centre for Quantum Computer Technology, Macquarie University,\\ Sydney, New South Wales 2109, Australia}
\date{Jan 30, 2006}
\begin{abstract}
 We describe a new design for a q-wire with perfect transmission using a uniformly coupled Ising spin chain subject to global (homogeneously-applied) pulses. Besides allowing for perfect transport of single qubits, the design also yields the perfect  ``mirroring'' of multiply encoded qubits within the wire. We further utilise this global-pulse generated perfect mirror operation as a ``clock cycle'' to perform universal quantum computation on these multiply encoded qubits where the interior of the q-wire  serves as the quantum memory while the q-wire ends perform the quantum computation. We theoretically describe the operation of single and two-qubit quantum logic gates and show that only $N-1$ complete mirror cycles are required to execute  a quantum Fourier transform on $N$ qubits encoded within the q-wire. 
\end{abstract}

\maketitle
\indent The development of protocols for transmitting quantum states is a particularly important problem in quantum computation. The ability to produce q-wires would allow quantum information to be moved around within a quantum processor. 
In the initial work \cite{Bose2003, Lloyd2003}, the transport of quantum states through {\em unmodulated} spin chains was examined and less-than-perfect transport fidelities were found \cite{Bose2003, Giovannetti2005, Subrahmanyam2004, Plenio2004, De2005a, Christandl2004, Christandl2005}. This is due to the dispersion of the quantum information along the chain \cite{Osborne2004}. Much work has since ensued searching for perfect q-wire transport schemes and briefly we can categorise these into:  (1) if the nearest-neighbour couplings between systems comprising the q-wire are set to very specific values 
\cite{Christandl2004, Nikolopoulos2004,  Yung2005, Christandl2005, Karbach2005}, one can achieve perfect transport. (2)  One can achieve near perfect transport by encoding the quantum information into low-dispersion wavepackets, or by encoding/decoding via conditional quantum logic across multiple q-wires 
\cite{Osborne2004, Haselgrove2004, Giovannetti2005, Burgarth2005b, Burgarth2005, Burgarth2005a}. (3) Use `gapped systems', where the q-wire ends are only weakly coupled to a strongly inter-coupled interior of the q-wire \cite{Plenio2005, Li2005, Wojcik2005}, to achieve near perfect transport. (4) Other possibilities include teleportation of the quantum information along the q-wire by measurements \cite{Barjaktarevic2005}, encoding into soliton-like excitations \cite{Khaneja2002},  or use quantum cellular automata concepts  \cite{Benjamin2002, Brennen2003}.  Besides the transport of single qubits, of more interest is the capability of the q-wire to transport entire qubit registers via `quantum mirror wires' \cite{Albanese2004, Karbach2005}. Here an unknown multi-qubit quantum state,  when encoded at one end of the wire is transmitted to the other end, but in reverse order, $\rho_{j_1j_2\cdots j_N}^{i_1i_2\cdots i_N}\in{\cal H}^1\otimes{\cal H}^2\otimes\cdots {\cal H}^ N\rightarrow \tilde{\rho}=\rho^{i_Ni_{N-1}\cdots i_1}_{j_N\cdots j_1}$.  Experimental proposals for q-wires include Josephson junction arrays \cite{Romito2005}, molecular magnet wires \cite{Blundell2004}, Quantum Nano-Electromechanical systems \cite{Eisert2004}, and  tunnel-coupled electronic quantum dots \cite{Nikolopoulos2004}. 

As well as demonstrating that globally addressed q-wires can yield perfect qubit transport and perfect multi-qubit mirroring we will also show that they can be used to execute universal quantum computation. We achieve this via a combination of the application of selective local unitaries on the ends of the q-wire and homogenous local unitaries (HLUs \cite{Masanes2002}), (or global pulses), on the entire q-wire. The use of HLUs alone to perform quantum computation has been examined by a number of authors \cite{Lloyd2003, Benjamin2000a, Benjamin2001, Twamley2003, Raussendorf2005}. In all but the last of these, the application of HLUs alone is not sufficient to implement universal quantum computation and some structuring of the q-wire is typically required, e.g. two or three types of cells in the q-wire. Our hybrid approach using HLUs and end-system selective addressing has a number of benefits over pure HLU computing. We require no structuring of the q-wire while the use of robust composite pulses \cite{Jones2003,McHugh2005}, can greatly reduce the effects of any static variations in the inter-system coupling strengths.  Finally, to our knowledge, no fault tolerant quantum error correction scheme has been found for pure HLU quantum computation. It is our hope that  such a scheme might be more feasible in our hybrid design.
It may be that such q-wires  could comprise both the computational and communication resources within a quantum processor and possibly lead to greater simplifications in the required technology. 
 \begin{figure}
\includegraphics[height=9.5cm,width=7cm]{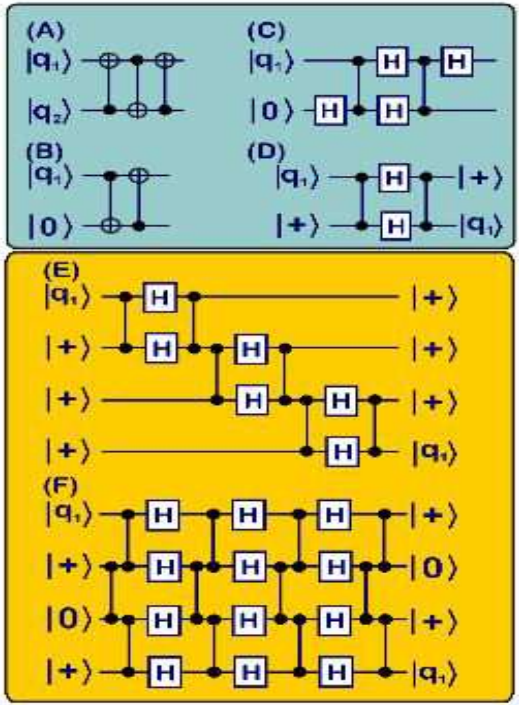}
\caption{We derive the network for globally controlled, perfect state transfer via several relatively simple steps. (A) a SWAP-gate, (B) same but with only $|q_1\rangle$ to SWAP. Using   $CNOT^{a,b} = H^b CZ^{a,b} H^b$, where $CZ$ is the $\pi/2$ phase gate, we arrive at (C), and recoding the known input state $|0\rangle\rightarrow |+\rangle$, we have (D). (E) Chaining  these operations together to perfectly transport an unknown state. (F) The input qubits $|q_1q_2\cdots q_n\rangle=|++\cdots +\rangle$, are obtained via the input state $|+0\cdots +\rangle$, and homogenous applications of Hadamard and $CZ$. \label{fig:fig1}}
\end{figure}

{\it State Transfer and Ising Interactions:-}
The simplest approach to state transfer in a q-wire is to simply swap qubits in neighbouring locations, repeating the process on alternating pairs of qubits until the desired state has reached the end of the q-wire, at which point no further swaps are performed. Building on this idea we follow the steps outlined in Figure 1 to arrive at the circuit (F), which transports unknown state $|q_1\rangle$  
using simultaneously applied Hadamard operations $\overline{H}
\equiv \prod_{j=1}^N \, \otimes H^j$, and controlled phase operations $\overline{CZ}\equiv \prod_{j=1}^{N-1}\, \otimes CZ^{j,j+1}$. Complete transport through a $N$-system q-wire with the initial state $|q_1+0+0\cdots\rangle$, requires the application of $\overline{CZ}\cdot(\overline{H}\cdot\overline{CZ})^{N-1}$ global operations.  From (F) it would appear that such transport will require the very particular initial state  $|q_1q_2q_3\cdots q_N\rangle=|q_1+0\cdots\rangle$, where $q_a$, $a=2,..,N$ are the $|+(0)\rangle$, pure states alternately. However this is not the case as we show below and any initial state of these other systems will suffice (even completely mixed states).

The execution of (F) in Figure \ref{fig:fig1} requires the application of the {\em global} pulses $\overline{H}$ and $\overline{CZ}$.  We have assumed that the interior q-wire systems are identical and $\overline{H}$ is generated via the single qubit operations on the degenerate interior systems with selective application of Hadamards simultaneously on the end systems (Note: typically interior and end systems will possess different resonant frequencies due to their difference in neighbor interactions). To execute $\overline{CZ}$,  we assume a uniform Ising interaction between all q-wire systems \cite{Ising_comment}.  Similar to the Hadamard global operation, we find that the execution of $\overline{CZ}$ consists of natural evolution under the Ising Hamiltonian together with single qubit operations which are uniform across the q-wire except at the end systems.  Allowing the uniform interaction $H_{Ising} = J \sum_{a=1}^{n-1} \sigma_z^a \sigma_z^{a+1}$, to run for $t = \frac{\pi \hbar}{4 J}$, we obtain $U_{Ising} = \exp(-i\frac{\pi}{4} \sum \sigma_z^{(a)} \sigma_z^{(a+1)})$. We can expand an individual $CZ^{a,b}= \frac{1}{2}(I + \sigma_z^a + \sigma_z^b - \sigma_z^a\sigma_z^b)$, and using this we can expand the full $\overline{CZ} = \prod_{a=1}^{N-1} {CZ}^{a,a+1} =\prod_{a=1}^{N-1} \frac{1}{2}(I + \sigma_z^a + \sigma_z^{a+1} - \sigma_z^a\sigma_z^{a+1})$. The generating Hamiltonian for this transformation is $H = \hbar g \sum_{a=1}^{n-1} \frac{1-\sigma_z^{(a)}}{2}\frac{1-\sigma_z^{(a+1)}}{2}$. This can be expanded and using $U_{Ising}$, we see that 
\begin{eqnarray}
\overline{CZ} & = & \exp(-i\frac{\pi}{4}(\sigma_z^{(1)}+ \sigma_z^{(N)}))(\prod \exp(i\frac{\pi}{2}\sigma_z^{(a)})) U_{Ising} \nonumber\\
& = & R_z^{(1)}(\frac{\pi}{4})R_z^{(N)}(\frac{\pi}{4})(\prod R_z^{(a)}(-\frac{\pi}{2}))U_{Ising},\label{Uising}
\end{eqnarray} where $R_z^{(a)}(\theta)$ is a $z-$rotation, performed on the system at position $a$. The single qubit operations in (\ref{Uising}) consists of a  $-\pi/2$ homogenous $z-$rotation on each q-wire system except for the end systems which have additional $\pi/4$ $z-$rotations (via selective pulses). Since these $z-$rotations commute with $U_{Ising}$, we can choose to execute them either before or after the Ising interaction.  A magnetic field along the $z-$axis could be used to perform the rotations while the Ising interaction is running, however this may not be optimal and we can choose to wrap these $z-$rotations in with the following $\overline{H}$ global operation (except for the last application of the $\overline{CZ}$, where these rotations must either be executed or passed on to the next computational element following the q-wire transmission). The evolution in (F) consists of repetitions of $\overline{H}\cdot\overline{CZ}$. Setting
$\overline{CZ}=\prod_{a=1}^N\,e^{i\theta_a\sigma_z^{(a)}}\,U_{Ising}$
where $\theta_a=\pi/4,\;\;a=1,N$, or $\theta_a=\pi/2,\;\;a\ne1,N$, and noting that $H \sigma_z H=\sigma_x$, and $HH=\mathbb{I}$, we have
\[
\overline{H}\cdot\overline{CZ}=
\left( \prod_{a=1}^N\,e^{i\theta_a\sigma_x^{a}}\right)\overline{H}U_{Ising}\;\;.
\]
We now use $-iH=\exp(-i\pi \sigma_x)\exp(-i\pi/2\sigma_y)$, to obtain
\[
\overline{H}\cdot \overline{CZ}=\left((-i)^N \prod_{a=1}^N\,e^{i\chi_a\sigma_x^{(a)}}e^{-i\pi/2\sigma_y^{(a)}}\right)U_{Ising}
\]
where $\chi_a=-3\pi/4,\;\;a=1,N$ and $\chi_a=-\pi/2,\;\;a\ne1,N$. Thus the combination of a homogenous application of $CZ$ gates on neighbouring sites on the wire, together with an ensuing application of local Hadamard gates on all sites can be generated via the standard Ising interaction following by `bang-bang' type homogenous local operations on each qubit (apart from two $\sigma_x$, operations applied at the end sites - which we assume can be selectively addressed apart from the bulk of the wire). Thus the perfect transport circuit of Figure  \ref{fig:fig1} (F), requires only global addressing of the the  q-wire, an Ising interaction which is uniform along the q-wire and selective manipulation of the q-wire end systems.   \\[1ex]

  \begin{figure}
\includegraphics[width=8cm]{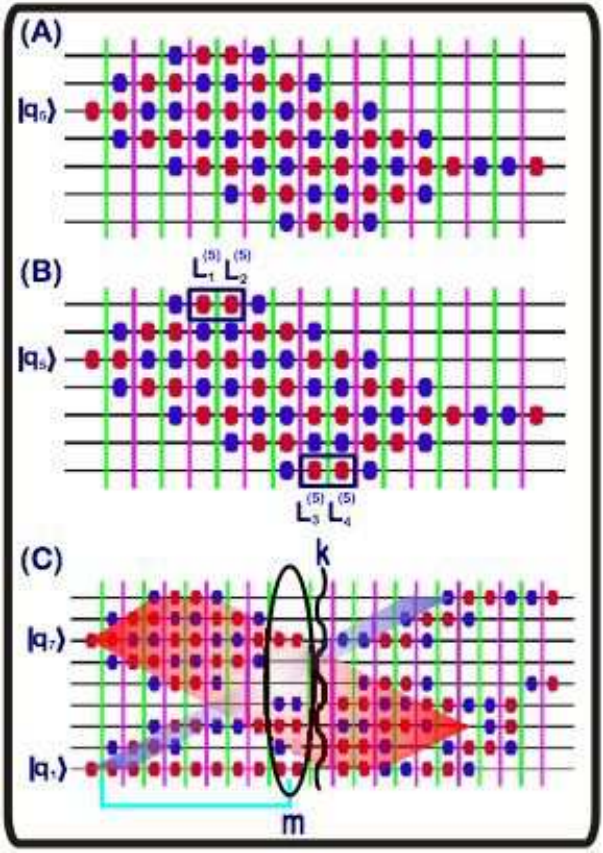}
\caption{Illustration of mirror transport, method of executing single and two qubit gates. (A) Mirror transport of the local unitary operations  $X$ (red), and  $Z$ (blue), acting on $|q_5\rangle$, through an $N=7$ length chain via the application of  $\mathcal{S}\equiv U^{Mirror}=(\overline{H}\cdot\overline{CZ})^{N+1}$. The  global pulses are $\overline{CZ}\equiv\prod^{N-1}_{i=1}CZ^{i,i+1}$ (green vertical bars), and $\overline{H}\equiv \prod_{i=1}^N H^i$ (purple vertical bars). 
(B)  To execute single qubit gates on a pattern (e.g. on $|q_4\rangle$), we apply  
 apply $\sigma_z$ operations on the end-systems of the q-wire at the times noted $L^{(5)}_1,..,L^{(5)}_4$.
 (C) Illustration of two-qubit gate between control $|q^c\rangle=|q_1\rangle$, and target $|q^t\rangle=|q_7\rangle$. Colored regions Blue (Red)  depict the separate transport patterns for an initial $X$ operation on $|q_1\rangle$( $|q_7\rangle$). Underlying pattern depicts the simultaneous transport of both initial qubits where we have set the $|q_1\rangle$ trap on during the indicated period (shown in powder blue). This traps a component of the $|q_1\rangle$ pattern along the bottom edge of the graph (C) until it reaches time-step $m$, where the trap is turned off. At $m$,  the  pattern from $|q_7\rangle$, begins to impact the trapped pattern and when this occurs we apply $\overline{CZ}\cdot\overline{CZ}$ to the entire q-wire  but lift off the trapping of the end site for a short time during this global pulse to yield a $CZ[\theta]=diag(1,1,1,e^{i\theta})$, or a controlled phase gate. At time $k$ we can either reverse the temporal order of the global operations to return to the initial state but where now $|\tilde{q}_1\rangle\otimes|\tilde{q}_7\rangle=CZ[\theta]|q_1\rangle\otimes|q_7\rangle$, or  we can continue forward, keeping the $|q_1\rangle$ pattern trapped to execute a number of controlled phase gates targeting any qubit pattern which impacts that trapped $|\tilde{q}_1\rangle$.
 \label{fig:fig2}}
\end{figure}

{\it Perfect Quantum Mirrors:-}
 The above perfect transport seems to depend having the particular initial state $|q_1+0\cdots\rangle$. Although this state allowed us to easily derive the transport circuit in Figure \ref{fig:fig1} (F), it is not necessary as the total operation $\mathcal{S}\equiv(\overline{H}\cdot\overline{CZ})^{N+1}$, constitutes a perfect quantum mirror which reverses the spatial location of any quantum information encoded on the q-wire. To see this we take, with no loss in generality, the initial state of an $N$-system q-wire to be in a pure product state  with $|\psi\rangle_{init}\sim \cdots \otimes |q_k\rangle\otimes$, where any pure state of the $k^{th}$ system can be expressed as $|q_k\rangle=\alpha_k|0\rangle_k+\beta_k|1\rangle_k=(\alpha_k+\beta_k\sigma_x^{(k)})|0\rangle_k$. To prove mirror transport it suffices to prove mirror transport of the initial state operations, i.e. $\mathcal{S}\sigma_x^{(k)}=\sigma_x^{(N-k+1)}\mathcal{S}$, (and similarly for $\sigma_z^{(a)}$). To prove these identities we make use of the following rules for propagating these operators through the global operations $\CZ$ and $\H$:
\[\CZ\sigma_z^{(a)}=\sigma_z^{(a)}\CZ\;\;,\;\;
\CZ\sigma_x^{(1)}=\sigma_x^{(1)}\sigma_z^{(2)}\CZ\;\;,\]\vspace{-.7cm}
\begin{eqnarray}
\CZ\sigma_x^{(N)}&=&\sigma_z^{(N-1)}\sigma_x^{(N)}\CZ\;\;,\nonumber\\
\CZ\sigma_x^{(a)}&=&\sigma_z^{(a-1)}\sigma_x^{(a)}\sigma_z^{(a+1)}\CZ\;\;,\nonumber\\
\H\sigma_z^{(a)}&=&\sigma_x^{(a)}\H\;\;.\label{rules}\\ \nonumber
\end{eqnarray}
Using these rules one can follow the propagation of $\sigma_x^{(a)}$ (or $\sigma_z^{(a)}$), through the global operations, e.g. $(\overline{H}\cdot\overline{CZ})^{2}\sigma_x^{(5)}= \sigma_x^{(3)}\sigma_z^{(4)}\sigma_x^{(5)}\sigma_z^{(6)}\sigma_x^{(7)}(\overline{H}\cdot\overline{CZ})^{2}$. The propagation can be more easily understood through a graphical representation (see Figure \ref{fig:fig2}(A)).  Using these rules and the graphical representation one can show
\begin{equation}
\mathcal{S}\sigma_x^{(a)}=\sigma_x^{(N-a+1)}\mathcal{S}\;\;,\qquad
\mathcal{S}\sigma_z^{(a)}=\sigma_z^{(N-a+1)}\mathcal{S}\;\;.
\end{equation}
We see from Figure \ref{fig:fig2}(A) that the propagation typically undergoes a period of expansion until the pattern hits the nearest q-wire end. It then continues to expand in the other direction while remaining ``stuck'' at the end it has impacted. Following two applications of $\overline{H}\cdot\overline{CZ}$ after impact  the pattern reflects off this nearest wire end and then the process of impact, sticking and reflection repeats off the other end of the q-wire. Following $N+1$ applications of clocking operation, $\overline{H}\cdot\overline{CZ}$, the initial product state of the q-wire undergoes a perfect spatial inversion about the wire's midpoint and consequently the inversion of any initial state of the q-wire occurs after a full {\em cycle} of $\mathcal{S}=(\overline{H}\cdot\overline{CZ})^{N+1}$ operations. 

The construction of perfect quantum mirror transport using only global operations, uniform Ising interactions and end-system selective manipulation, could be ideally suited to transport entire qubit registers throughout a quantum processor and may need only modest technological developments to become possible in the near future in a variety of physical implementations.

\section*{Q-Wire Quantum Computing}\vspace{-.25cm}
{\it Single Qubit Gates:-}
\indent Besides quantum transport we show below how a q-wire can be used to perform universal quantum computation on a register of qubits encoded within the wire. To achieve this we make full use of the capability to separately manipulate the end systems of the q-wire. As mentioned above, such a capability should be the typical case as the end systems are naturally differentiated from the internal q-wire systems. We refer back to Figure \ref{fig:fig2}(A) and the space-time patterns of qubit operations. We note that the pattern resulting from a single qubit operation acting on the initial state impacts a horizontal edge of this pattern in a series of four cells. From now we assume that the initial qubit register is padded by $|+\rangle$ states, i.e. $|\psi\rangle_{init}=|q_1\,+\,q_2\,+q_3\,+\cdots\rangle$. This is to ensure that in the following the required edge operations (at the top and bottom rows of the space-time pattern), do not interfere with each other. To execute universal quantum logic we must demonstrate single qubit operations and two qubit conditional operations. To achieve the former, the execution of a general qubit rotation, $U^{(a)}(\alpha,\beta,\gamma)\equiv R_z(\alpha)R_y(\beta)R_z(\gamma)$, on any qubit  $q_a$, we use three full mirror cycles of $\mathcal{S}$. During the first cycle, to execute $R_z(\gamma)$, on $q_a$, we apply this single qubit operation on an edge at the impact points labelled $L^{(a)}_i$, in Figure \ref{fig:fig2}(B). Upon the execution of the first $\mathcal{S}$, where now the qubit register is spatially reversed along the q-wire we apply the global operation $\overline{H}_y\equiv\prod_{a=1}^N\,\otimes H_y$, where $H_y\equiv R_x^{(a)}(\pi/2)$, on the entire q-wire. We then apply the second round of $\mathcal{S}$, and during this apply $R_z(\beta)$, at any of the edge locations $L^{(a)}_i$. Once the second $\mathcal{S}$ is completed we again apply $\overline{H}_y$ to the q-wire and using $H_y R_z^{(a)}(\theta)H_y=R_y^{(a)}(\theta)$, we see that this second round yields $R_y(\beta)$ on $q_a$. In the third round of $\mathcal{S}$, we again apply $R_z(\alpha)$, at any of the locations $L^{(a)}_i$ to arrive at $U^{(a)}(\alpha,\beta,\gamma)$. With our use of buffer states the data operator patterns do not overlap on the top and bottom edges of Figure \ref{fig:fig2}(B) and consequently we are able to execute $\prod_{a=1}^N \otimes U^{(a)}(\alpha_a,\beta_a,\gamma_a)$, i.e. arbitrary single qubit operations on all qubits encoded within the q-wire, using three rounds of $\mathcal{S}$, using edge operations and global $\overline{H}_y$.

%

{\it Two Qubit Gates:-}
 To execute two qubit gates we utilise the end-system control to apply decoupling pulses selectively to either end-system (one can use pulses such as those described in \cite{Stollsteimer2001}, but more simply one can apply a $R_x(\pi/2)$, pulse to the end spin mid-way through the Ising gate to average out the Ising interaction completely), or use selective pulses to move an end-system qubit to an ``off-line'' storage memory using techniques such as those recently demonstrated in a Nitrogen-Vacancy-${}^{13}$C coupled system \cite{Jelezko2004}. By decoupling off an end-system we artificially shorten the q-wire and by continuing to apply the global operations $\overline{H}\cdot\overline{CZ}$, (while omitting the Hadamard on the decoupled end site), we can cycle the remaining qubits within this shortened q-wire. Since we are unable to apply the local $R_z^{(N-1)} (-\pi/4)$ necessary to complete $\overline{CZ}$ on the shortened q-wire, an unwanted $R_z^{(N-1)} (\pi/4)$ is introduced each time we $\overline{CZ}$.

 To execute a control-phase gate on qubit  $|q^t\rangle$, controlled by the state of $|q^c\rangle$, both encoded in different spatial sites within the q-wire, we wait until the $X$-pattern from $|q^c\rangle$ impacts an end of the q-wire whereupon we apply a decoupling pulse sequence to trap this $X$ pattern at the end of the q-wire. The target qubit pattern will cycle forward and will reach a configuration where it commences to impact the trapped $X$ pattern from $|q^c\rangle$ (see Figure \ref{fig:fig2}(C)). Then, instead of $\overline{CZ}\cdot\overline{H}\cdot\overline{CZ}$, we apply the Ising interaction for a time period $\tau_{2\pi}$, to yield, $\overline{CZ}\cdot\overline{CZ}$, on all systems within the q-wire, while lifting the decoupling of the end-system for a time $\tau_\theta=\tau_{2\pi}*(\theta/2\pi)$, during this global operation. The result of this is to execute the identity operation on all qubits within the q-wire bar the end-system qubit and its immediate neighbour which suffer a control-$\theta$ operation and along with a $R_z^{(N-1)} ((\pi-\theta)/4)$. All the operations up to this point, except for the control-$\theta$ and its $R_z^{(N-1)} ((\pi-\theta)/4)$, will later be reversed, and so we can tolerate the additional rotations as long as the commute with the control-$\theta$ operator between the last two qubits.

As rotations introduced at the edge of the chain propogate away from the edge, only the $R_z^{(N-1)}(-\pi/4)$ introduced with the final $\overline{CZ}$ prior to the controlled-$\theta$ gate does not commute with it. To overcome this, we apply a global $R_z (\pi/4)$, cancelling the rotation on qubit $N-1$, but introducing rotations on qubits $1...(N-2)$. Clearly, these commute with the controlled-$\theta$ gate between the last two qubits.

 We now wish to undo everything except the controlled-$\theta$ gate, to return to the initial state where now $|\tilde{q}^c\rangle\otimes|\tilde{q}^t\rangle=R_z^{(t)}((\pi-\theta)/4)CZ[\theta]|q^c\rangle\otimes|q^t\rangle$, and where $CZ[\theta]=diag(1,1,1,e^{i\theta})$. This can be done by applying, in reverse order, the inverse of each gate used to reach this point. $\overline{H}$ and $\overline{CZ}$ are, obviously, their own inverses, and $R_z(-\pi/4)$ is the inverse of the global $R_z(\pi/4)$. It is important to note, however, that $\overline{CZ_D}$, the result of applying $\overline{CZ}$ with the end-system qubit decoupled, is not its own inverse. This requires us to use $\overline{CZ_D'} = R_z^{(1)}(-\frac{\pi}{4})(\prod R_z^{(a)}(\frac{\pi}{2}))U_{Ising}^3$ instead of $\overline{CZ_D}$ when reversing the trapping sequence. This will remove any unwanted rotations introduced by not correcting the extra $R_z^{(N-1)} (\pi/4)$ caused at each $\overline{CZ_D}$.

However, more usefully, instead of reversing the temporal order of the global pulses we reverse only as far as the global $R_z (\pi/4)$ before continuing forward in the cycling evolution of the q-wire patterns, while still keeping the end system trapped to repeat the execution of a further $CZ[\theta_2]$, on another target qubit. Continuing in this fashion we can execute $CZ[\theta_1, \theta_2, \cdots, \theta_{n-1}]|q^c, q^t_1,q^t_2, \cdots q^t_{n-1}\rangle$, for a q-wire encoding $n$ qubits. At the end of a full cycle we release the trap and return the control qubit back to its original spatial location in the q-wire. This multi-target 2-qubit gate can provide significant savings when it comes to executing more complicated quantum circuits such as the quantum fast Fourier transform.

{\it Quantum Fast Fourier Transform:-}
The quantum Fourier can be written as
\[
QFT = H_N W^{N-1} H_{N-1} \ldots W^2 H_2 W^1 H_1
\]
\noindent where
\[
W^x = \ket{0_x}\bra{0_x} \otimes I + \ket{1_x}\bra{1_x} \otimes \prod_{j \neq x} W_j
\]
\noindent with $W_j = R_j(\pi / 2^{j-x})$ for $j>x$ and $W_j = I_j$ otherwise. Clearly, $W^x$ is composed of $N-x$ individual controlled phase (CPHASE) gates. Thus the quantum Fourier transform can be constructed using $(N-2)(N-1)/2$ controlled phase gates and N Hadamard gates. As has been shown earlier, arbitrary CPHASEs ($CZ[\theta]$), can be easily implemented in this scheme. Furthermore, all the CPHASE gates controlled by a particular qubit can be performed in at most a single mirroring cycle of the system. Thus, each $W_x$ term takes only a single mirror cycle. The corresponding Hadamard gate can also be performed during this cycle, reducing the time required to perform a QFT to only N-1 mirror cycles of the system.

We have demonstrated a protocol where one can implement perfect quantum transport, quantum mirroring and quantum computation through an Ising-coupled q-wire where one requires only the additional capability of executing global pulses on all systems within the q-wire and selective pulses on the ends of the q-wire. 
We have found how to execute the quantum Fourier transform of $N$ encoded qubits in $N-1$ mirror cycles. By using robust pulse techniques the protocol should be very insensitive to substantial variations in the inter-system couplings within the q-wire. Potential experimental demonstrations might include chains of superconducting systems or Nitrogen-Vacancy defects in diamond. The relatively simple design of the globally controlled q-wire may also have applications beyond quantum information processing such as in the design of new types of engineered quantum materials.

\bibliography{Chain_compute}

\begin{thebibliography}{37}
\expandafter\ifx\csname natexlab\endcsname\relax\def\natexlab#1{#1}\fi
\expandafter\ifx\csname bibnamefont\endcsname\relax
  \def\bibnamefont#1{#1}\fi
\expandafter\ifx\csname bibfnamefont\endcsname\relax
  \def\bibfnamefont#1{#1}\fi
\expandafter\ifx\csname citenamefont\endcsname\relax
  \def\citenamefont#1{#1}\fi
\expandafter\ifx\csname url\endcsname\relax
  \def\url#1{\texttt{#1}}\fi
\expandafter\ifx\csname urlprefix\endcsname\relax\def\urlprefix{URL }\fi
\providecommand{\bibinfo}[2]{#2}
\providecommand{\eprint}[2][]{\url{#2}}

\bibitem[{\citenamefont{Bose}(2003)}]{Bose2003}
\bibinfo{author}{\bibfnamefont{S.}~\bibnamefont{Bose}}, \bibinfo{journal}{Phys.
  Rev. Lett.} \textbf{\bibinfo{volume}{91}}, \bibinfo{pages}{207901}
  (\bibinfo{year}{2003}).

\bibitem[{\citenamefont{Lloyd}(2003)}]{Lloyd2003}
\bibinfo{author}{\bibfnamefont{S.}~\bibnamefont{Lloyd}},
  \bibinfo{journal}{Phys. Rev. Lett.} \textbf{\bibinfo{volume}{90}},
  \bibinfo{pages}{167902} (\bibinfo{year}{2003}).

\bibitem[{\citenamefont{Giovannetti and Fazio}(2005)}]{Giovannetti2005}
\bibinfo{author}{\bibfnamefont{V.}~\bibnamefont{Giovannetti}} \bibnamefont{and}
  \bibinfo{author}{\bibfnamefont{R.}~\bibnamefont{Fazio}},
  \bibinfo{journal}{Phys. Rev. A} \textbf{\bibinfo{volume}{71}},
  \bibinfo{pages}{032314} (\bibinfo{year}{2005}).

\bibitem[{\citenamefont{Subrahmanyam}(2004)}]{Subrahmanyam2004}
\bibinfo{author}{\bibfnamefont{V.}~\bibnamefont{Subrahmanyam}},
  \bibinfo{journal}{Phys. Rev. A} \textbf{\bibinfo{volume}{69}},
  \bibinfo{pages}{034304} (\bibinfo{year}{2004}).

\bibitem[{\citenamefont{Plenio et~al.}(2004)\citenamefont{Plenio, Hartley, and
  Eisert}}]{Plenio2004}
\bibinfo{author}{\bibfnamefont{M.~B.} \bibnamefont{Plenio}},
  \bibinfo{author}{\bibfnamefont{J.}~\bibnamefont{Hartley}}, \bibnamefont{and}
  \bibinfo{author}{\bibfnamefont{J.}~\bibnamefont{Eisert}},
  \bibinfo{journal}{New J. Phys.} \textbf{\bibinfo{volume}{6}},
  \bibinfo{pages}{36} (\bibinfo{year}{2004}).

\bibitem[{\citenamefont{De~Chiara et~al.}(2005)\citenamefont{De~Chiara,
  Rossini, Montangero, and Fazio}}]{De2005a}
\bibinfo{author}{\bibfnamefont{G.}~\bibnamefont{De~Chiara}},
  \bibinfo{author}{\bibfnamefont{D.}~\bibnamefont{Rossini}},
  \bibinfo{author}{\bibfnamefont{S.}~\bibnamefont{Montangero}},
  \bibnamefont{and} \bibinfo{author}{\bibfnamefont{R.}~\bibnamefont{Fazio}},
  \bibinfo{journal}{Phys. Rev. A} \textbf{\bibinfo{volume}{72}},
  \bibinfo{pages}{012323} (\bibinfo{year}{2005}).

\bibitem[{\citenamefont{Christandl et~al.}(2004)\citenamefont{Christandl,
  Datta, Ekert, and Landahl}}]{Christandl2004}
\bibinfo{author}{\bibfnamefont{M.}~\bibnamefont{Christandl}},
  \bibinfo{author}{\bibfnamefont{N.}~\bibnamefont{Datta}},
  \bibinfo{author}{\bibfnamefont{A.}~\bibnamefont{Ekert}}, \bibnamefont{and}
  \bibinfo{author}{\bibfnamefont{A.~J.} \bibnamefont{Landahl}},
  \bibinfo{journal}{Phys. Rev. Lett.} \textbf{\bibinfo{volume}{92}},
  \bibinfo{pages}{187902} (\bibinfo{year}{2004}).

\bibitem[{\citenamefont{Christandl et~al.}(2005)\citenamefont{Christandl,
  Datta, Dorlas, Ekert, Kay, and Landahl}}]{Christandl2005}
\bibinfo{author}{\bibfnamefont{M.}~\bibnamefont{Christandl}},
  \bibinfo{author}{\bibfnamefont{N.}~\bibnamefont{Datta}},
  \bibinfo{author}{\bibfnamefont{T.~C.} \bibnamefont{Dorlas}},
  \bibinfo{author}{\bibfnamefont{A.}~\bibnamefont{Ekert}},
  \bibinfo{author}{\bibfnamefont{A.}~\bibnamefont{Kay}}, \bibnamefont{and}
  \bibinfo{author}{\bibfnamefont{A.~J.} \bibnamefont{Landahl}},
  \bibinfo{journal}{Phys. Rev. A} \textbf{\bibinfo{volume}{71}},
  \bibinfo{pages}{032312} (\bibinfo{year}{2005}).

\bibitem[{\citenamefont{Osborne and Linden}(2004)}]{Osborne2004}
\bibinfo{author}{\bibfnamefont{T.~J.} \bibnamefont{Osborne}} \bibnamefont{and}
  \bibinfo{author}{\bibfnamefont{N.}~\bibnamefont{Linden}},
  \bibinfo{journal}{Phys. Rev. A} \textbf{\bibinfo{volume}{69}},
  \bibinfo{pages}{052315} (\bibinfo{year}{2004}).

\bibitem[{\citenamefont{Nikolopoulos et~al.}(2004)\citenamefont{Nikolopoulos,
  D.Petrosyan, and Lambropoulos}}]{Nikolopoulos2004}
\bibinfo{author}{\bibfnamefont{G.~M.} \bibnamefont{Nikolopoulos}},
  \bibinfo{author}{\bibnamefont{D.Petrosyan}}, \bibnamefont{and}
  \bibinfo{author}{\bibfnamefont{P.}~\bibnamefont{Lambropoulos}},
  \bibinfo{journal}{Condens. Matter} \textbf{\bibinfo{volume}{16}},
  \bibinfo{pages}{4991} (\bibinfo{year}{2004}).

\bibitem[{\citenamefont{Yung and Bose}(2005)}]{Yung2005}
\bibinfo{author}{\bibfnamefont{M.~H.} \bibnamefont{Yung}} \bibnamefont{and}
  \bibinfo{author}{\bibfnamefont{S.}~\bibnamefont{Bose}},
  \bibinfo{journal}{Phys. Rev. A} \textbf{\bibinfo{volume}{71}},
  \bibinfo{pages}{032310} (\bibinfo{year}{2005}).

\bibitem[{\citenamefont{Karbach and Stolze}(2005)}]{Karbach2005}
\bibinfo{author}{\bibfnamefont{P.}~\bibnamefont{Karbach}} \bibnamefont{and}
  \bibinfo{author}{\bibfnamefont{J.}~\bibnamefont{Stolze}},
  \bibinfo{journal}{Phys. Rev. A} \textbf{\bibinfo{volume}{72}},
  \bibinfo{pages}{030301} (\bibinfo{year}{2005}), \bibinfo{note}{e-print
  quant-ph/0501007}.

\bibitem[{\citenamefont{Haselgrove}(2004)}]{Haselgrove2004}
\bibinfo{author}{\bibfnamefont{H.~L.} \bibnamefont{Haselgrove}}
  (\bibinfo{year}{2004}), \bibinfo{note}{e-print quant-ph/0404152}.

\bibitem[{\citenamefont{Burgarth and Bose}(2005{\natexlab{a}})}]{Burgarth2005b}
\bibinfo{author}{\bibfnamefont{D.}~\bibnamefont{Burgarth}} \bibnamefont{and}
  \bibinfo{author}{\bibfnamefont{S.}~\bibnamefont{Bose}},
  \bibinfo{journal}{Phys. Rev. A} \textbf{\bibinfo{volume}{71}},
  \bibinfo{pages}{052315} (\bibinfo{year}{2005}{\natexlab{a}}).

\bibitem[{\citenamefont{Burgarth et~al.}(2005)\citenamefont{Burgarth,
  Giovannetti, and Bose}}]{Burgarth2005}
\bibinfo{author}{\bibfnamefont{D.}~\bibnamefont{Burgarth}},
  \bibinfo{author}{\bibfnamefont{V.}~\bibnamefont{Giovannetti}},
  \bibnamefont{and} \bibinfo{author}{\bibfnamefont{S.}~\bibnamefont{Bose}},
  \bibinfo{journal}{J. Phys. A: Math. Gen.} \textbf{\bibinfo{volume}{38}},
  \bibinfo{pages}{6793} (\bibinfo{year}{2005}).

\bibitem[{\citenamefont{Burgarth and Bose}(2005{\natexlab{b}})}]{Burgarth2005a}
\bibinfo{author}{\bibfnamefont{D.}~\bibnamefont{Burgarth}} \bibnamefont{and}
  \bibinfo{author}{\bibfnamefont{S.}~\bibnamefont{Bose}}, \bibinfo{journal}{New
  J. Phys.} \textbf{\bibinfo{volume}{7}}, \bibinfo{pages}{135}
  (\bibinfo{year}{2005}{\natexlab{b}}).

\bibitem[{\citenamefont{Plenio and Semião}(2005)}]{Plenio2005}
\bibinfo{author}{\bibfnamefont{M.}~\bibnamefont{Plenio}} \bibnamefont{and}
  \bibinfo{author}{\bibfnamefont{F.~L.} \bibnamefont{Semião}},
  \bibinfo{journal}{New J. Phys.} \textbf{\bibinfo{volume}{7}},
  \bibinfo{pages}{73} (\bibinfo{year}{2005}).

\bibitem[{\citenamefont{Li et~al.}(2005)\citenamefont{Li, Shi, Chen, Song, and
  Sun}}]{Li2005}
\bibinfo{author}{\bibfnamefont{Y.}~\bibnamefont{Li}},
  \bibinfo{author}{\bibfnamefont{T.}~\bibnamefont{Shi}},
  \bibinfo{author}{\bibfnamefont{B.}~\bibnamefont{Chen}},
  \bibinfo{author}{\bibfnamefont{Z.}~\bibnamefont{Song}}, \bibnamefont{and}
  \bibinfo{author}{\bibfnamefont{C.~P.} \bibnamefont{Sun}},
  \bibinfo{journal}{Phys. Rev. A} \textbf{\bibinfo{volume}{71}},
  \bibinfo{pages}{022301} (\bibinfo{year}{2005}).

\bibitem[{\citenamefont{Wojcik et~al.}(2005)\citenamefont{Wojcik, Luczak,
  Kurzynski, Grudka, Gdala, and Bednarska}}]{Wojcik2005}
\bibinfo{author}{\bibfnamefont{A.}~\bibnamefont{Wojcik}},
  \bibinfo{author}{\bibfnamefont{T.}~\bibnamefont{Luczak}},
  \bibinfo{author}{\bibfnamefont{P.}~\bibnamefont{Kurzynski}},
  \bibinfo{author}{\bibfnamefont{A.}~\bibnamefont{Grudka}},
  \bibinfo{author}{\bibfnamefont{T.}~\bibnamefont{Gdala}}, \bibnamefont{and}
  \bibinfo{author}{\bibfnamefont{M.}~\bibnamefont{Bednarska}}
  (\bibinfo{year}{2005}), \bibinfo{note}{e-print quant-ph/0505097}.

\bibitem[{\citenamefont{Barjaktarevic et~al.}(2005)\citenamefont{Barjaktarevic,
  Links, Milburn, and McKenzie}}]{Barjaktarevic2005}
\bibinfo{author}{\bibfnamefont{J.~P.} \bibnamefont{Barjaktarevic}},
  \bibinfo{author}{\bibfnamefont{J.}~\bibnamefont{Links}},
  \bibinfo{author}{\bibfnamefont{G.}~\bibnamefont{Milburn}}, \bibnamefont{and}
  \bibinfo{author}{\bibfnamefont{R.~H.} \bibnamefont{McKenzie}}
  (\bibinfo{year}{2005}), \bibinfo{note}{e-print quant-ph/0501180}.

\bibitem[{\citenamefont{Khaneja and Glaser}(2002)}]{Khaneja2002}
\bibinfo{author}{\bibfnamefont{N.}~\bibnamefont{Khaneja}} \bibnamefont{and}
  \bibinfo{author}{\bibfnamefont{S.~J.} \bibnamefont{Glaser}},
  \bibinfo{journal}{Phys. Rev. A} \textbf{\bibinfo{volume}{66}},
  \bibinfo{pages}{060301} (\bibinfo{year}{2002}).

\bibitem[{\citenamefont{Benjamin}(2002)}]{Benjamin2002}
\bibinfo{author}{\bibfnamefont{S.~C.} \bibnamefont{Benjamin}},
  \bibinfo{journal}{Phys. Rev. Lett.} \textbf{\bibinfo{volume}{88}},
  \bibinfo{pages}{107904} (\bibinfo{year}{2002}).

\bibitem[{\citenamefont{Brennen and Williams}(2003)}]{Brennen2003}
\bibinfo{author}{\bibfnamefont{G.~K.} \bibnamefont{Brennen}} \bibnamefont{and}
  \bibinfo{author}{\bibfnamefont{J.~E.} \bibnamefont{Williams}},
  \bibinfo{journal}{Phys. Rev. A} \textbf{\bibinfo{volume}{68}},
  \bibinfo{pages}{042311} (\bibinfo{year}{2003}).

\bibitem[{\citenamefont{Albanese et~al.}(2004)\citenamefont{Albanese,
  Christandl, Datta, and Ekert}}]{Albanese2004}
\bibinfo{author}{\bibfnamefont{C.}~\bibnamefont{Albanese}},
  \bibinfo{author}{\bibfnamefont{M.}~\bibnamefont{Christandl}},
  \bibinfo{author}{\bibfnamefont{N.}~\bibnamefont{Datta}}, \bibnamefont{and}
  \bibinfo{author}{\bibfnamefont{A.}~\bibnamefont{Ekert}},
  \bibinfo{journal}{Phys. Rev. Lett.} \textbf{\bibinfo{volume}{93}},
  \bibinfo{pages}{230502} (\bibinfo{year}{2004}).

\bibitem[{\citenamefont{Romito et~al.}(2005)\citenamefont{Romito, Fazio, and
  Bruder}}]{Romito2005}
\bibinfo{author}{\bibfnamefont{A.}~\bibnamefont{Romito}},
  \bibinfo{author}{\bibfnamefont{R.}~\bibnamefont{Fazio}}, \bibnamefont{and}
  \bibinfo{author}{\bibfnamefont{C.}~\bibnamefont{Bruder}},
  \bibinfo{journal}{Phys. Rev. B} \textbf{\bibinfo{volume}{71}},
  \bibinfo{pages}{100501} (\bibinfo{year}{2005}).

\bibitem[{\citenamefont{Blundell and Pratt}(2004)}]{Blundell2004}
\bibinfo{author}{\bibfnamefont{S.~J.} \bibnamefont{Blundell}} \bibnamefont{and}
  \bibinfo{author}{\bibfnamefont{F.~L.} \bibnamefont{Pratt}},
  \bibinfo{journal}{J. Phys.: Condens. Matter} \textbf{\bibinfo{volume}{16}},
  \bibinfo{pages}{R771} (\bibinfo{year}{2004}).

\bibitem[{\citenamefont{Eisert et~al.}(2004)\citenamefont{Eisert, Plenio, Bose,
  and Hartley}}]{Eisert2004}
\bibinfo{author}{\bibfnamefont{J.}~\bibnamefont{Eisert}},
  \bibinfo{author}{\bibfnamefont{M.}~\bibnamefont{Plenio}},
  \bibinfo{author}{\bibfnamefont{S.}~\bibnamefont{Bose}}, \bibnamefont{and}
  \bibinfo{author}{\bibfnamefont{J.}~\bibnamefont{Hartley}},
  \bibinfo{journal}{Phys. Rev. Lett.} \textbf{\bibinfo{volume}{93}},
  \bibinfo{pages}{190402} (\bibinfo{year}{2004}).

\bibitem[{\citenamefont{Masanes et~al.}(2002)\citenamefont{Masanes, Vidal, and
  Latorre}}]{Masanes2002}
\bibinfo{author}{\bibfnamefont{L.}~\bibnamefont{Masanes}},
  \bibinfo{author}{\bibfnamefont{G.}~\bibnamefont{Vidal}}, \bibnamefont{and}
  \bibinfo{author}{\bibfnamefont{J.~I.} \bibnamefont{Latorre}},
  \bibinfo{journal}{Quant. Inf \& Comp.} \textbf{\bibinfo{volume}{2}},
  \bibinfo{pages}{285} (\bibinfo{year}{2002}).

\bibitem[{\citenamefont{Benjamin}(2000)}]{Benjamin2000a}
\bibinfo{author}{\bibfnamefont{S.~C.} \bibnamefont{Benjamin}},
  \bibinfo{journal}{Phys. Rev. A} \textbf{\bibinfo{volume}{6102}},
  \bibinfo{pages}{020301} (\bibinfo{year}{2000}).

\bibitem[{\citenamefont{Benjamin}(2001)}]{Benjamin2001}
\bibinfo{author}{\bibfnamefont{S.~C.} \bibnamefont{Benjamin}},
  \bibinfo{journal}{Phys. Rev. A} \textbf{\bibinfo{volume}{6405}},
  \bibinfo{pages}{054303} (\bibinfo{year}{2001}).

\bibitem[{\citenamefont{Twamley}(2003)}]{Twamley2003}
\bibinfo{author}{\bibfnamefont{J.}~\bibnamefont{Twamley}},
  \bibinfo{journal}{Phys. Rev. A} \textbf{\bibinfo{volume}{67}},
  \bibinfo{pages}{052318} (\bibinfo{year}{2003}).

\bibitem[{\citenamefont{Raussendorf}(2005)}]{Raussendorf2005}
\bibinfo{author}{\bibfnamefont{R.}~\bibnamefont{Raussendorf}},
  \bibinfo{journal}{Phys. Rev. A} \textbf{\bibinfo{volume}{72}},
  \bibinfo{pages}{052301} (\bibinfo{year}{2005}).

\bibitem[{\citenamefont{Jones}(2003)}]{Jones2003}
\bibinfo{author}{\bibfnamefont{J.~A.} \bibnamefont{Jones}},
  \bibinfo{journal}{Phys. Rev. A} \textbf{\bibinfo{volume}{67}},
  \bibinfo{pages}{012317} (\bibinfo{year}{2003}).

\bibitem[{\citenamefont{McHugh}(2005)}]{McHugh2005}
\bibinfo{author}{\bibfnamefont{D.}~\bibnamefont{McHugh}}, Ph.D. thesis,
  \bibinfo{school}{National University of Ireland, Maynooth}
  (\bibinfo{year}{2005}).

\bibitem[{Isi()}]{Ising_comment}
\bibinfo{note}{We suspect that one can use interactions other than the Ising
  interaction to implement $\bar{CZ}$.}

\bibitem[{\citenamefont{Stollsteimer and Mahler}(2001)}]{Stollsteimer2001}
\bibinfo{author}{\bibfnamefont{M.}~\bibnamefont{Stollsteimer}}
  \bibnamefont{and} \bibinfo{author}{\bibfnamefont{G.}~\bibnamefont{Mahler}},
  \bibinfo{journal}{Phys. Rev. A} \textbf{\bibinfo{volume}{64}},
  \bibinfo{pages}{052301} (\bibinfo{year}{2001}).

\bibitem[{\citenamefont{Jelezko et~al.}(2004)\citenamefont{Jelezko, Gaebel,
  Popa, Domhan, Gruber, and Wrachtrup}}]{Jelezko2004}
\bibinfo{author}{\bibfnamefont{F.}~\bibnamefont{Jelezko}},
  \bibinfo{author}{\bibfnamefont{T.}~\bibnamefont{Gaebel}},
  \bibinfo{author}{\bibfnamefont{I.}~\bibnamefont{Popa}},
  \bibinfo{author}{\bibfnamefont{M.}~\bibnamefont{Domhan}},
  \bibinfo{author}{\bibfnamefont{A.}~\bibnamefont{Gruber}}, \bibnamefont{and}
  \bibinfo{author}{\bibfnamefont{J.}~\bibnamefont{Wrachtrup}},
  \bibinfo{journal}{Phys. Rev. Lett.} \textbf{\bibinfo{volume}{93}},
  \bibinfo{pages}{130501} (\bibinfo{year}{2004}).

\end{thebibliography}

\small{Correspondence and requests for material should be sent to
J. T. (e-mail: jtwamley@ics.mq.edu.au).}

\end{document}